\documentclass{PoS}

\title{SUSY Higgs bosons and beyond}

\ShortTitle{SUSY Higgs bosons and beyond}

\author{Marcela Carena\\
        Fermilab National Accelerator Laboratory and University of Chicago, USA\\
        E-mail: \email{carena@fnal.gov}}

\author{Eduardo Ponton\\
        Columbia University, USA\\
        E-mail: \email{eponton@phys.columbia.edu}}

\author{\speaker{Jos\'e Zurita} \\%
        University of Zurich, Switzerland\\
        E-mail: \email{jzurita@physik.uzh.ch}}

\abstract{We consider extensions of the Minimal Supersymmetric Standard Model (MSSM) where the extra degrees of freedom interact weakly with the Higgs sector. These models allow to relax the tension between the lower bound on the lightest CP even Higgs mass from direct LEP searches and the theoretical upper bound of the MSSM. We study the beyond MSSM (BMSSM) effects via an effective field-theory approach, assuming that the MSSM is valid up to a heavy physics scale $M$. We compute the masses, couplings and branching fractions of the Higgs sector, including all the relevant corrections up to order $1/M^2$.  We find that the collider phenomenology can be greatly different with respect to both the SM and the MSSM.}

\FullConference{XVIII International Workshop on Deep-Inelastic Scattering and
Related Subjects\\
                 April 19 -23, 2010\\
                 Convitto della Calza, Firenze, Italy}

\begin{document}

\section{Introduction}
\label{sec:intro}

In the Minimal Supersymmetric Standard Model (MSSM), the mass of the lightest CP-even Higgs has an experimental lower bound of 90 GeV \cite{Schael:2006cr}, and a theoretical upper bound of around 130 GeV \cite{hmass}. It is well known that the theoretical constraint can be lifted by introducing new degrees of freedom, and along those lines many particular extensions of the MSSM have been studied \cite{raising}. 
In order to be as general as possible, we will treat the extra physics with an effective field theory approach, as was performed already in \cite{Brignole:2003cm}, and 
\cite{Dine:2007xi}. Our goal is to study the collider phenomenology of these models.

The Higgs sector of the MSSM consists of two doublets, 
$H_u$ and $H_d$, each of them with its own vacuum expectation value ($v_u$ and $v_d$). After the spontaneous breakdown of the gauge symmetry, one has five degrees of freedom left, that correspond to two scalars ($h$ and $H$), a pseudoscalar ($A$), and a pair of charged Higgs bosons ($H^{\pm}$). The Higgs potential has seven quartic couplings, $\lambda_{1-7}$,
instead of only one as in the Standard Model (SM). 
At tree level, the whole sector depends only on two parameters, that are usually chosen to be $m_A$ and $\tan \beta=v_u / v_d $. and one finds $m_h \leq m_Z | \cos(2 \beta) |$. 
Taking radiative corrections into account the upper bound can reach values of around 130 GeV, where the specific value will depend upon the supersymmetric soft parameters.
In Section~\ref{sec:review} we review the basic setup of the BMSSM Higgs sectors (see \cite{Carena:2009gx} for details) while in Section~\ref{sec:results} we show a selected sample of the results obtained in \cite{Carena:2010cs}.
\section{BMSSM Higgs sectors}
\label{sec:review}
In this setup, the extra-MSSM 
degrees of freedom are treated with an effective field theory approach, which is valid up to a 
scale $M$. At first order in the $1/M$ expansion there are only two new operators \cite{Dine:2007xi}, namely
\begin{equation}
\label{eq:dim5}
W=\mu H_u H_d + \frac{\omega_1}{2M} (1 + \alpha_1 X) (H_u H_d)^2 \,
\end{equation}
where $X=m_S \theta^2$ is a spurion superfield, $m_S$ is the SUSY breaking scale in the heavy sector, and $\mu$ stands for the MSSM $\mu$-term. We assume that $\alpha_1$ and $\omega_1$ are order one coefficients. In this work we will pick $\mu=200~{\rm GeV} = m_S$ and $M=1~{\rm TeV}$.
The amazingly simple structure of Eq.~(\ref{eq:dim5}) triggered the detailed study of the Higgs potential \cite{hdo}. Consequences for dark matter \cite{hdo:dm}, cosmology \cite{hdo:cosmo} and electroweak baryogenesis \cite{hdo:baryo} have also been explored. 

At order $1/M^2$, one has many more new operators \cite{Carena:2009gx,Antoniadis:2009rn}. 
We will argue that these operators play an important role. The reason is that at tree level $\lambda_{1-4}$ are very small (of the order of the electroweak gauge couplings) and do not get $1/M$ corrections, while
$\lambda_{5-7}$ are zero at tree level but not at order $1/M$. Therefore a full dimension six operator analysis has to be performed.

We have computed the corrections to $\lambda_{1-7}$ from dimension five and six operators, adding the 1-loop SUSY contributions from \cite{Carena:1995bx}.
The branching fractions for all the relevant decay modes of the MSSM Higgs bosons were obtained using a modified version of HDECAY \cite{Djouadi:1997yw}. In order to include the constraints coming from current collider data, we employ HiggsBounds \cite{Bechtle:2008jh}, and add
the bounds from LEP on charged Higgs bosons \cite{:2001xy} and the latest Tevatron data on $h \to WW$ \cite{Aaltonen:2010yv} and $H/h/A \to \tau \bar{\tau}$ \cite{:tau_inclusive} searches. 
We also impose the electroweak precision constraints 
and a convergence criteria for the $1/M$ expansion (see \cite{Carena:2009gx} for details).
 
\section{Collider Phenomenology}
\label{sec:results}
\begin{figure}[tp]
\begin{center}
\includegraphics[width=6.9cm]{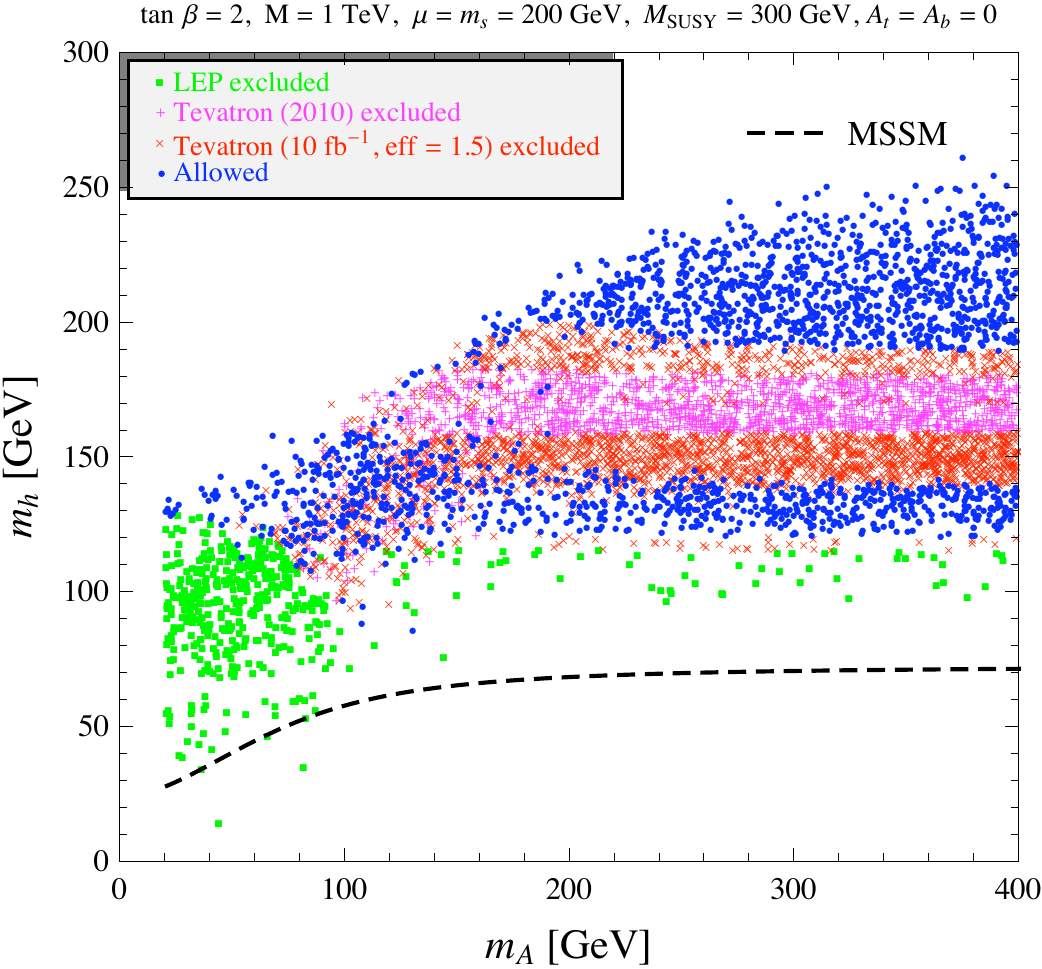}
\hspace{2mm}
\includegraphics[width=6.9cm]{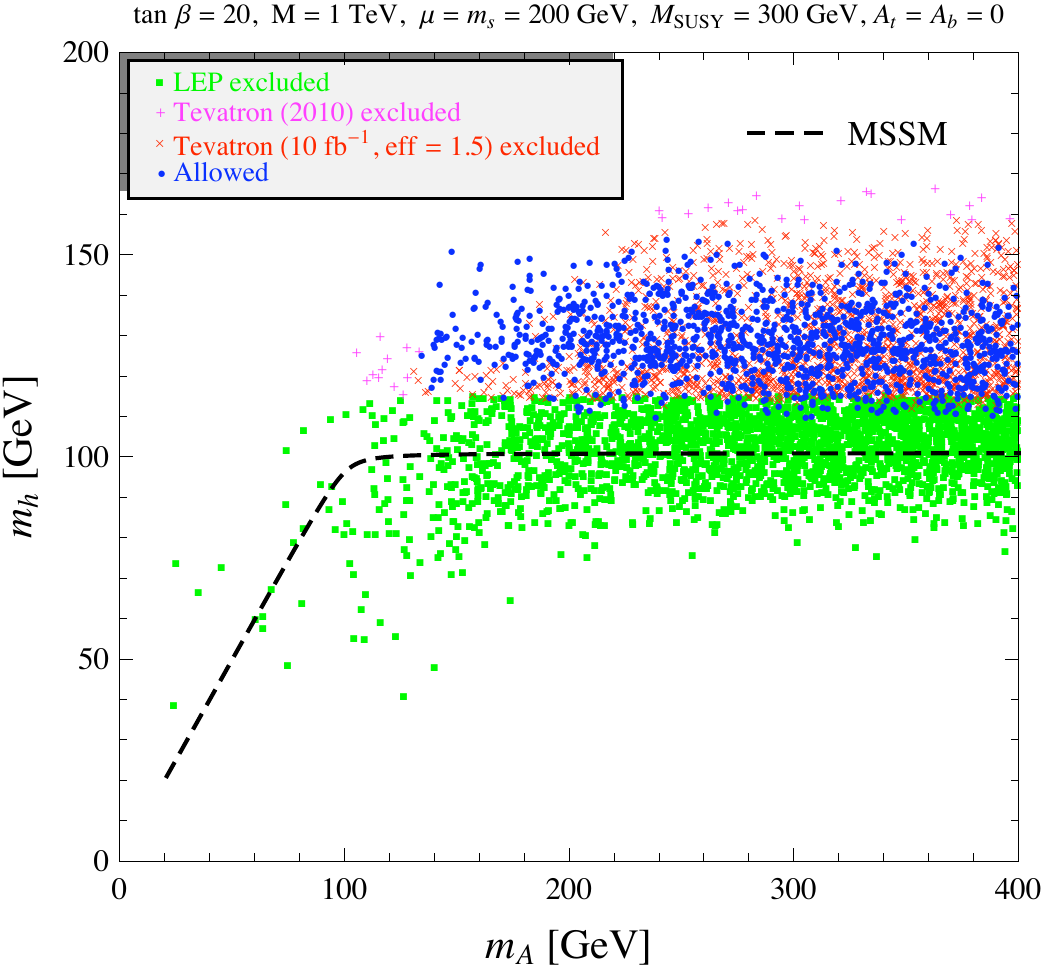}
\end{center}
\vspace{-0.5cm} 
\caption{\label{fig:mlvsma}{\em Lightest CP-even Higgs
boson mass as a function of $m_A$, for $\tan \beta=2$ (left panel) and
$\tan \beta=20$ (right panel).  We show the points excluded by LEP
(green), excluded by current Tevatron data (magenta) and the region
that will be probed by the Tevatron in the near future (red).  The
blue points are allowed by all the current experimental constraints.
The dashed line is the MSSM result for the given SUSY spectrum.  }}
\end{figure}
In Fig.~\ref{fig:mlvsma} we show $m_h$ as a function of $m_A$, for $\tan \beta=2$ and $20$. The green points are excluded by LEP, the magenta ones by current Tevatron data. The red ones refer to the scheduled run until 2011, where we have assumed a total integrated luminosity of 10 fb$^{-1}$ and a 50\% improvement in the efficiency of the $b \bar{b}$ and $WW$ channels \cite{Moriond_Fisher} (as was assumed in \cite{Draper:2009fh} for the MSSM analysis). 
The blue points are allowed by all current collider data.

We see that the effects on $m_h$ are larger for the low $\tan \beta$ regime, although for large values of $\tan \beta$ the effect is still not negligible: the values that can be obtained go easily  above the MSSM upper bound. For $\tan \beta=2$ we see blue, red and magenta stripes, telling us that the exclusion is mostly due to $h$ searches. It is worth noticing the presence of blue points with rather low values of $m_A$, that will give rise to situations where the decay modes $h/H \to AA$ might have order one branching fractions.

\begin{figure}[!t]
\begin{center}
\includegraphics[width=6.9cm]{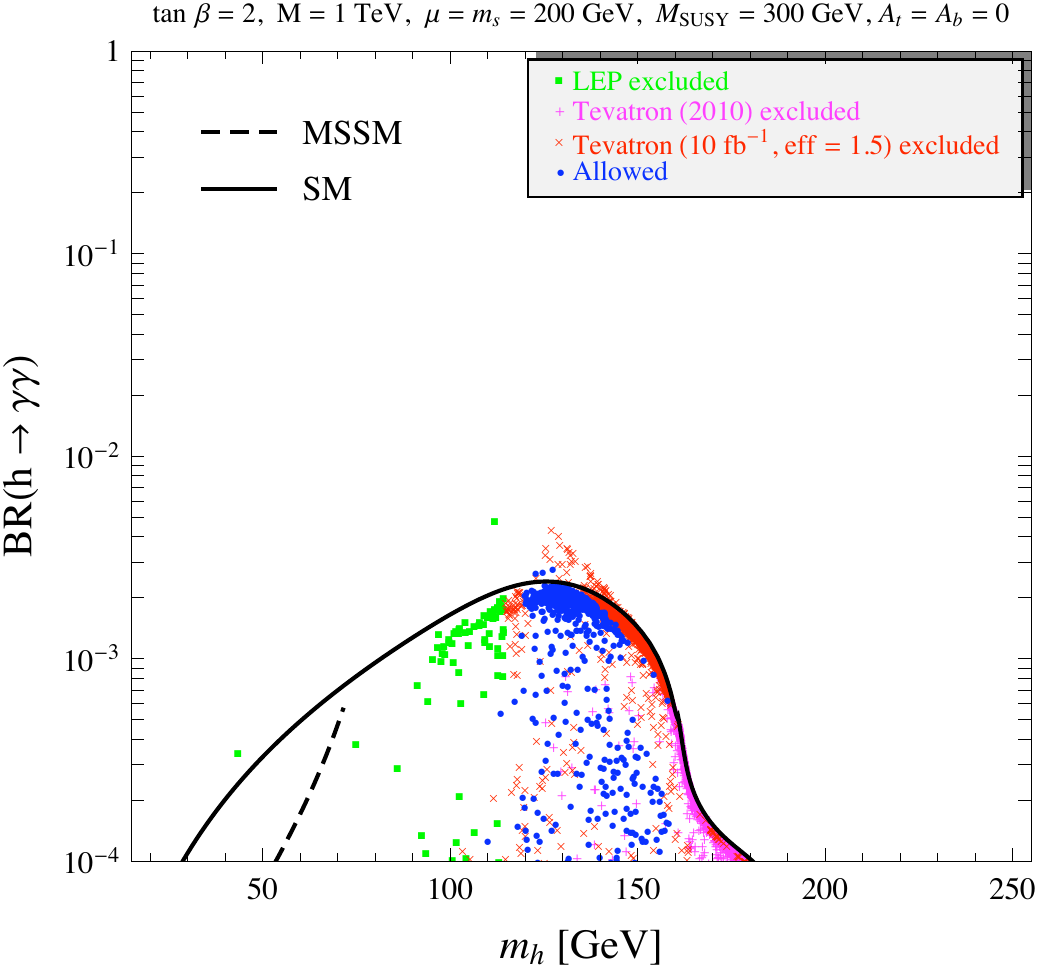}
\hspace{3mm}
\includegraphics[width=6.9cm]{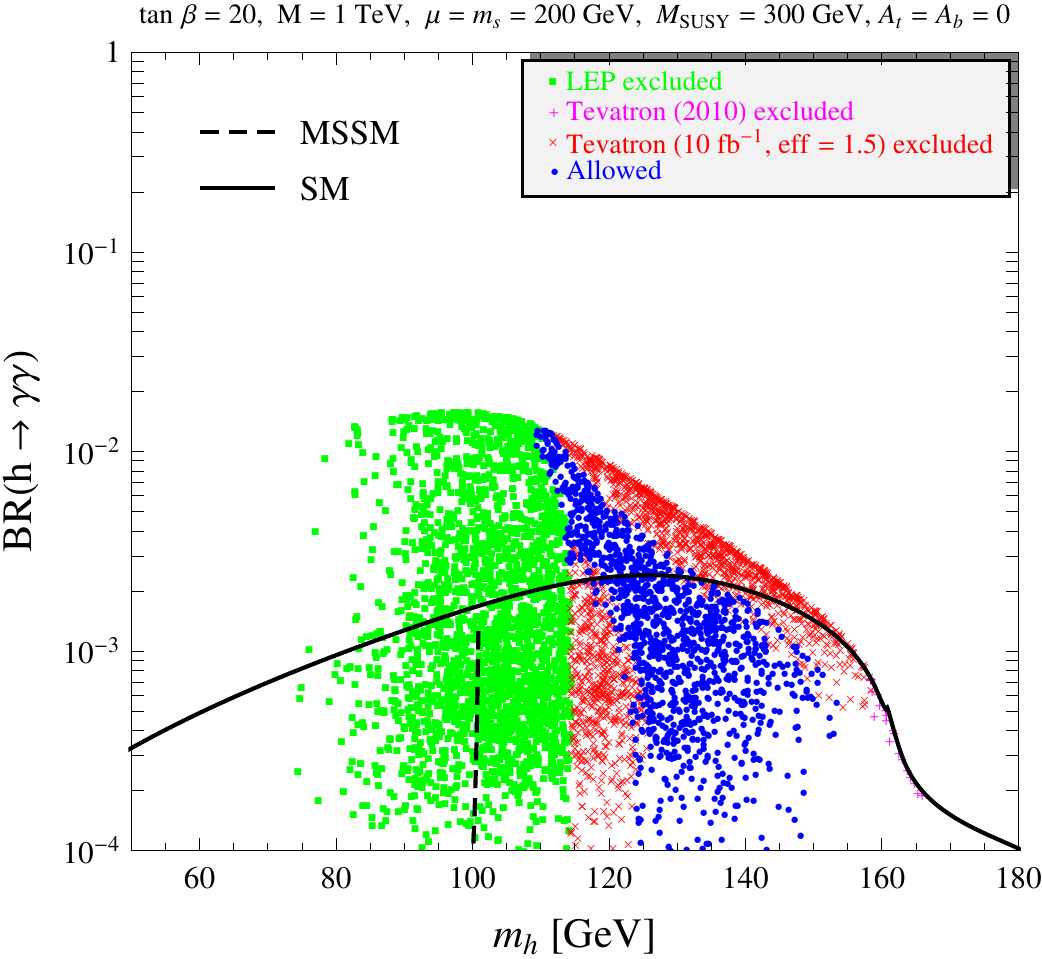}
\end{center}
\vspace{-0.5cm} 
\caption{\label{fig:htobgamtb2} {\em Branching fraction of $h \to \gamma \gamma$ for $\tan \beta = 2$ (left panel) and $\tan \beta=20$ (right panel). The color code is the same used in Fig.~1. The solid (dashed) line corresponds to the SM (MSSM) result.}}
\end{figure}

In Fig.~\ref{fig:htobgamtb2} we show the branching
fraction of $h \to \gamma \gamma$ as function of  $m_A$, for $\tan \beta=2$ and $20$.
From the left panel it is clear that in the low $\tan \beta$ regime, the branching fraction into the diphoton channel tends to be slightly suppressed w.r.t. SM. On the contrary, for large values of $\tan \beta$ one gets enhancements up to a factor of $8$ over the SM case. This is a very interesting feature not only for LHC searches, but also for the Tevatron: this factor has to be contrasted against
 the recent CDF \cite{:cdftwogammas} (D0 \cite{:d0twogammas}) analysis, that report an observed limit of $18.7-25.9$ $(11.9-28.3)$ in the $110-130~{\rm GeV}$ range, using $5.4~(4.2)$ fb$^{-1}$ of integrated luminosity. 
\section{Conclusions}
\label{sec:conclu}
In this work we have studied extensions of the MSSM with an effective-field theory approach up to the second order in the $1/M$ expansion, finding 
a very different phenomenology with respect to the MSSM.
The most striking feature is that a sizable rise of the lightest CP-even Higgs mass can be attained (specially for the low $\tan \beta$ regime), thus relaxing the tension between the theoretical upper bound and the exclusion from direct searches at LEP in the MSSM context. However, not only the Higgs spectrum, but also the couplings and the branching ratios can suffer important modifications w.r.t both the SM and the MSSM, thus leading to very interesting consequences for the collider phenomenology. As an example, we have shown the BR($h \to \gamma \gamma$), that in the large $\tan \beta$ regime can be sizably enhanced w.r.t SM, which is a very interesting feature for the forthcoming searches of Higgs bosons at both the Tevatron and the LHC.

\end{document}